\documentclass[slac_one]{revtex4}
\usepackage{graphicx}
\usepackage{fancyhdr}
\usepackage{subfigure}
\begin{document}

\newlength{\ogltemp}

\def\etmiss{\settowidth{\ogltemp}{\big\slash}%
\hspace{0.5mm}\big\slash\hspace{-1.4\ogltemp}{E_T}%
}
\def\ptmiss{\settowidth{\ogltemp}{\big\slash}%
\hspace{0.5mm}\big\slash\hspace{-1.4\ogltemp}{P_T}%
}
\def\vmet{\settowidth{\ogltemp}{\big\slash}%
\hspace{0.5mm}\big\slash\hspace{-1.4\ogltemp}{\vec{E}_T}%
}

\def\mel{\big\slash\hspace{-1.0ex}{\ell}}

\def\us{\char`\_}

\title{Search for SM Higgs Boson Produced in Association with a Z or a W
  Boson in events with $\etmiss$ and $b$-jets} 

%

\author{A. Apresyan (on behalf of the CDF collaboration)}
\affiliation{Purdue University, West Lafayette, IN 47907, USA}

\begin{abstract}
  We present a search for the Standard Model Higgs boson produced in
  association with a Z or a W boson, using data collected with the CDF
  II detector at the Tevatron accelerator. A scenario where the Z decays
  into neutrinos or leptons originating from the W-decay escape
  detection and the Higgs decays into a $b\overline{b}$ pair is
  considered. Therefore the expected signature is large missing
  transverse energy ($\etmiss$), no isolated leptons, and two b-jets.
  We present the preliminary results in this search using 2.1$fb^{-1}$
  of data collected by CDF.
\end{abstract}

\maketitle

\thispagestyle{fancy}


\section{\label{sec:Intro}INTRODUCTION} 
In the Higgs mechanism of the Standard Model, the fermions and weak
gauge bosons acquire mass via interaction with the Higgs field
\cite{higgspaper}.  Therefore the existence of this yet to be discovered
particle is the cornerstone of the Standard Model. The search for the
Higgs boson is one of the most active areas of research at the Tevatron.
The electroweak fits to SM parameters, performed including the latest
Tevatron top mass averaged measurements, point to the value
$m_H=84^{+34}_{-26}\,GeV/c^2$, which yields an upper limit of $m_H <
154\,GeV/c^2$\ at 95\% C.L. \cite{limits}. In the mass region below 135
GeV/$c^2$ the searches of Higgs boson production in association with a
vector boson provide the cleanest signatures due to the requirement of the
presence of decay products of the Z or W. The dominant decay mode of the
Higgs boson in this mass range is $H\rightarrow b\overline b$.

We present a search for the Higgs boson in events with no identified
leptons, large missing transverse energy ($\etmiss$) and two or three
jets, at least one of which originates from a $b$-quark. This channel is especially
challenging since the cross-sections of the background processes are
many orders of magnitude higher than the signal, even after requiring
the final state topology. If the Higgs boson mass is less than 135 GeV,
the associated production cross sections are of the order of 0.3--0.2
pb, which results in the signal to background ratio of about $1/20000$
after the trigger selections. It is thus necessary to employ
sophisticated event selection which reduces backgrounds to a more
manageable size.

\section{\label{sec:Data}{DATA SAMPLE AND EVENT SELECTION}}

We use data collected by the CDF experiment through August 2007, which
corresponds to 2.1 $fb^{-1}$ integrated luminosity. The events are
collected by CDF II detector with a trigger that selects events with
$\etmiss>25$ GeV at Level 1 at least two Level 2 clusters with
$E_{T}>10$ GeV and $\etmiss>35$ GeV at Level 3.

In the first step of the analysis, both the Monte Carlo simulation and
the CDF data events must pass quality cuts to ensure that the
possible beam and detector effects are removed from the data sample. The
standard CDF jet clustering algorithm is used with a jet
cone of radius 0.4. Jet energies are corrected for calorimeter
non-uniformity, non-linearity and energy loss in the un-instrumented
regions of calorimeter and energy coming from different $p \overline{p}$
interactions during the same bunch crossing.

The key element to distinguish the Higgs boson signal from background
events is the invariant mass of the $b\overline{b}$ system, which for
signal has a sharp peak around the Higgs boson mass, while the
backgrounds have smoother distributions. Obtaining the best possible
resolution of the $m_{b\overline{b}}$ is thus crucial for Higgs boson
searches, since it provides the most discriminating feature of the
signal. In order to improve the $m_{b\overline{b}}$ resolution, we
correct jet energies by reconstructing their four-momenta according to
an algorithm developed by H1 collaboration, using tracking information
\cite{H1}. This technique allows us to improve the dijet mass resolution
by around 10\%.

Traditionally the searches in this channel have analyzed events with two
jets, in order to keep the background rates low. For the first time at
CDF in this analysis we analyzed events with three jets in this channel.
The main motivation is to accept events where one of the {\em b} quarks
coming from the Higgs boson radiates a gluon, resulting in a three jet
event. In addition to that, we add acceptance to WH events where the
charged lepton coming from the W is reconstructed as a jet. The latter
case happens when the W decays to $e \nu$ and the electron fails the
isolation requirements, and is reconstructed as a jet; or when the $W$
decays to $\tau \nu$ and $\tau \to hadrons$. This is the first CDF
search to make use of events in $WH\rightarrow\tau\nu+b\overline{b}$
channel with $\tau$ identified as a jet. According to Monte-Carlo
simulations of WH events, around 30\% of three jet signal events
accepted in this analysis contain one jet matched to hadronic decays of
a $\tau$.

As a way to get a better estimate of the event true missing energy we
calculate the $\ptmiss$, which is defined as the magnitude of the
negative vectorial sum of charged particle track $p_T$'s. For true
$\etmiss$ events $\ptmiss$ is highly correlated with calorimetric
$\etmiss$, while for QCD events with mismeasured jets it is not. Thus,
$\ptmiss$ provides an additional handle to separate mismeasurements
from real $\etmiss$ events.

In order to enhance the signal purity, we require that at least one of
the two leading jets in the event is identified as a $b$-jet. Jets
originating from $b$-quarks can be identified ({\em ``$b$-tagged''}) by
exploiting the long lifetime of the $b$ hadrons by finding their decay
vertices, which are significantly displaced from the interaction point.
While the events with two $b$-tags provide the most sensitivity in this
analysis, the single-tagged sample adds about $10\%$ to the overall
sensitivity.

\section{\label{sec:BCKG}BACKGROUND ESTIMATION}

The contributions from the following background processes are considered
in this analysis: QCD multi-jet production, top quark pair and single
production, W or Z boson production with jets and diboson production
(WW,WZ,ZZ). The processes which yield events with $b$ or $c$ quarks,
were estimated using Monte-Carlo calculations.

The most significant background in this analysis is the QCD multijet
production, which has a production cross-section of several orders of magnitude
higher than signal. Although these processes generally do not have
intrinsic $\etmiss$, mismeasured jets can cause imbalance in the total
transverse energy, faking the signal if one of the jets is mis-tagged.
Furthermore, QCD b-quark pair production yields taggable jets and if one
$b$ undergoes a semi-leptonic decay large $\etmiss$. The QCD multijet
background is estimated from data, which allows us to estimate not only
the heavy flavor QCD production, but also processes with a light flavor
jet falsely tagged as a $b$-quark. The data driven technique enabled us
to considerably reduce the systematic uncertainty associated with the
QCD estimation.

In order to validate our ability to predict the sample composition we
test our model in three separate control region. All the important kinematic
variables and correlations between them are tested in these regions,
before analyzing the data in the signal region, i.e. we perform a
\emph{``blind''} analysis.

\section{\label{sec:SIGNAL}THE SEARCH FOR THE SIGNAL}

\begin{figure}[ht]
  \centering 
  \subfigure[QCD rejection
  NN]{\label{fig:QCDNN}\includegraphics[width=0.32\textwidth]{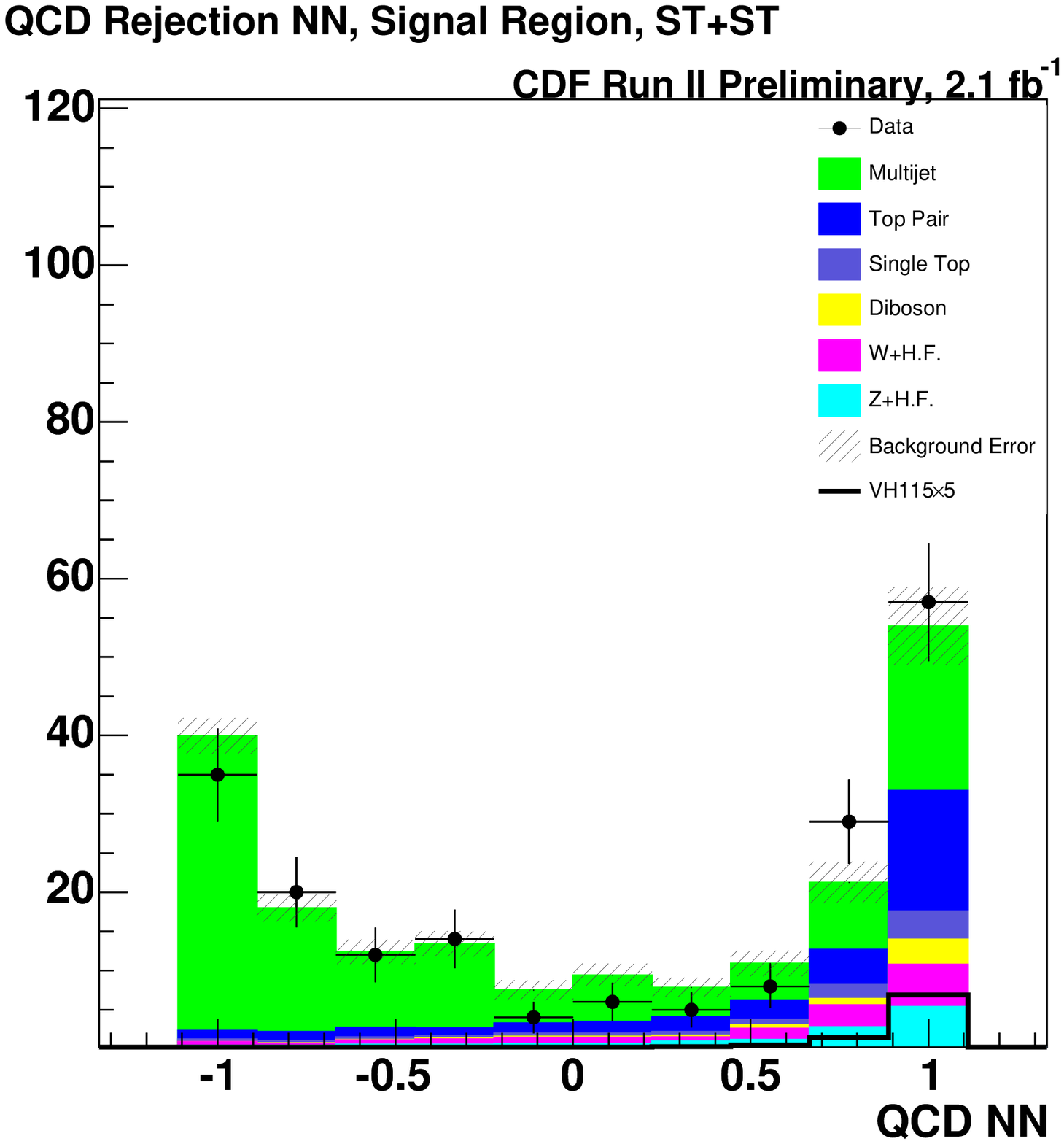}}
  \subfigure[Invariant mass of all
  jets]{\label{fig:mjjj}\includegraphics[width=0.32\textwidth]{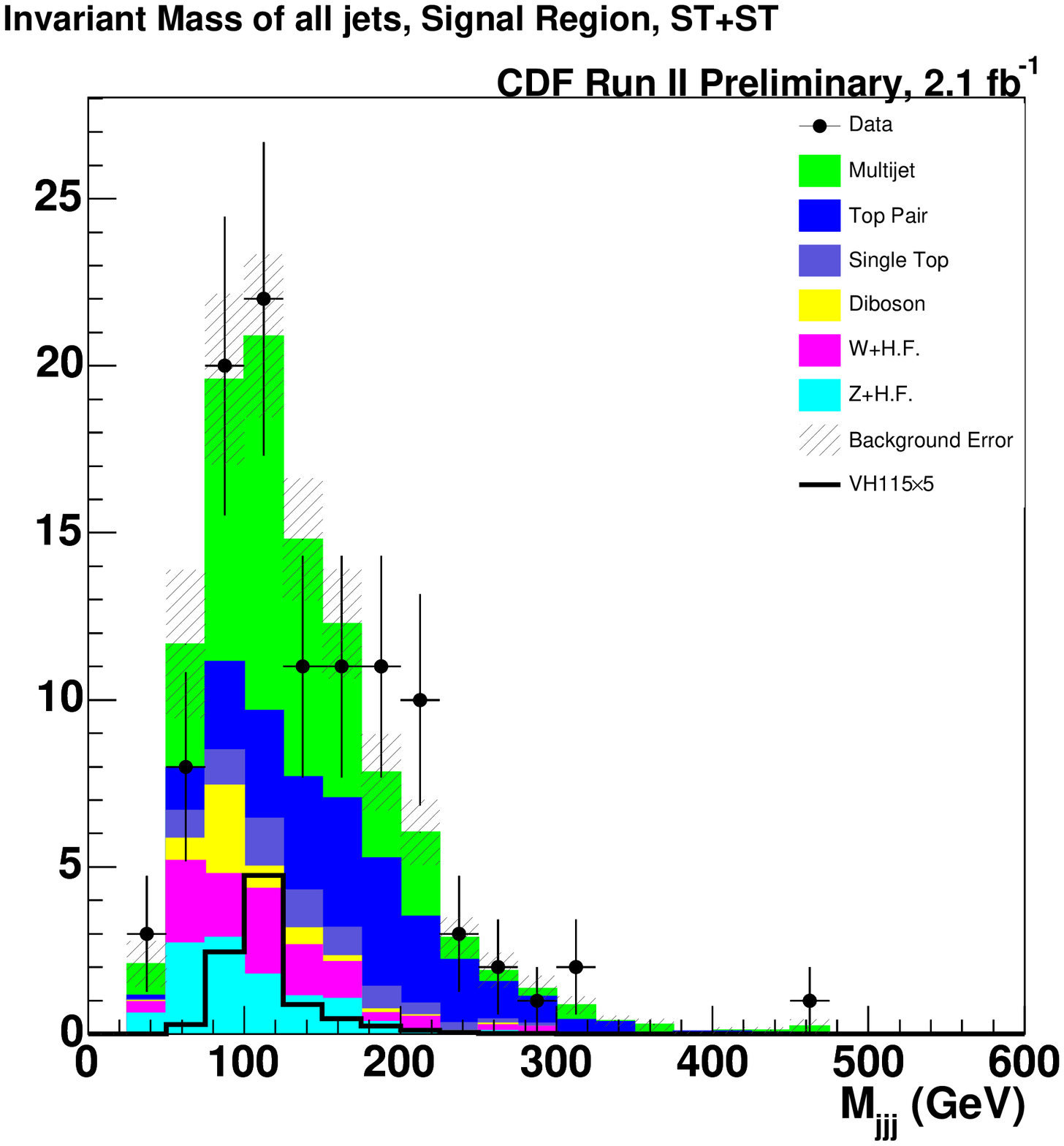}}
  \subfigure[Discriminant NN
  Output]{\label{fig:NN}\includegraphics[width=0.32\textwidth]{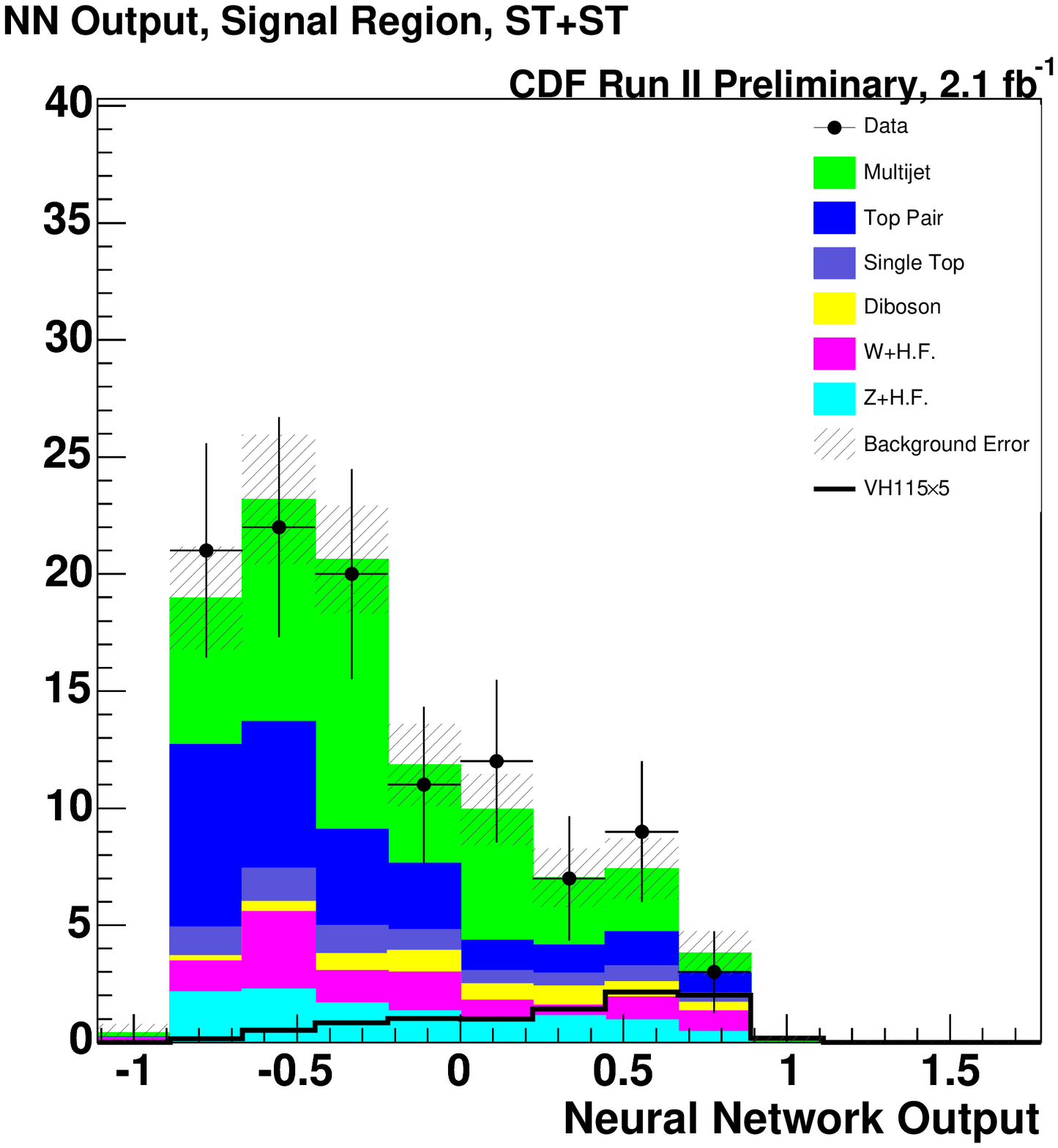}}
  \caption{\label{fig:outNNdisc} Comparisons of the background prediction and CDF observed data in double SecVTX tagged data in the Signal Region}
\end{figure}

We employ an Artificial Neural Network (NN) using Multi Layer Perceptron
\cite{tmva} to reject the majority of QCD mulitijet events, which are
the main source of background in this analysis. These processes have a
very different behavior with respect to the signal and the remaining
backgrounds, since they are mostly caused by instrumental effects. The
network is trained on a mixture of 50$\%$ WH and 50$\%$ ZH events
against QCD events. In order to achieve the maximum sensitivity, all the
events with the output of the QCD rejection NN below zero are rejected,
Fig.\ref{fig:QCDNN}. The remaining events constitute the Signal Region,
where we will search for a presence of the Higgs boson. This selection
requirement provides 95\% signal efficiency while rejecting 50\% of all
the backgrounds. By employing three exclusive tagging categories we
expect 7.3 signal events after final selections, assuming $m_H=115$
GeV/$c^2$. The expected signal yield is a factor of 2 greater than in
the past versions of the analysis, due to improved event selection
techniques and use of events in multiple tagging categories.

Once the signal region is defined, we train a discriminant NN to further
separate multijet and top pair production from signal, Fig.\ref{fig:NN}.
The discriminant NN uses seven input variables, including invariant mass
of all jets in the event, which is shown in Fig.\ref{fig:mjjj}. We scan
the distribution of the output of the discriminant NN for an excess of
observed data using binned likelihood technique. The likelihoods from
events with one or two $b$-tags are multiplied together.

\section{\label{sec:RESULTS}RESULTS}

Analyzing 2.1 $fb^{-1}$ of the CDF data we find a good agreement between
expected background predictions and the observed data. In the sample
with two $b$-tags, which provides the most sensitivity in this search,
we find 105 candidate events with an expectation of $103\pm 15$
background and 1.9 signal events, assuming $m_H=115$ GeV/$c^2$. Since no
significant excess is observed, we compute the expected upper limit for
the Higgs boson cross-section when the Higgs boson is produced in
association with a Z/W boson and decays to two $b$-quarks where Z decays
to neutrinos and W to leptons. We use the Bayesian likelihood method for
deriving the expected and observed upper limit at 95\% C.L. of 6.3 and
7.9 times the Standard Model cross section respectively for $m_H$ = 115
GeV/$c^2$ \cite{CDFMETBB08}. This result achieves an improvement of 30\%
in sensitivity with respect to the previous version of the search in the
same channel, and provides one of the most stringent limits on Higgs
boson production cross-section among various Tevatron searches.


\begin{thebibliography}{99} 
\bibitem{higgspaper} P.W.Higgs, Phys. Rev. Lett., \textbf{13}, 508
  (1964)
\bibitem{limits} The LEP Electroweak Working Group: \url{http://lepewwg.web.cern.ch/LEPEWWG/}
\bibitem{H1} H1 Collaboration, C. Adloff et al., Z. Phys. C74 (1997)
  221.
\bibitem{tmva} A.~Hocker {\it et al.}, ``TMVA: Toolkit for multivariate
  data analysis,'' arXiv:physics/0703039.
\bibitem{CDFMETBB08} CDF Collaboration, ``Search for the Standard Model
  Higgs boson in the MET plus jets sample'', CDF9483

\end{thebibliography}
\end{document}